\documentclass[prl,aps,twocolumn,floatfix,superscriptaddress]{revtex4-2}
\usepackage{color}
\usepackage[colorlinks,linktocpage,linkcolor=blue]{hyperref}

\usepackage{eso-pic,calc}
\usepackage{graphicx}
\usepackage{amsmath, amsthm, amssymb, mathrsfs}
\usepackage{epsfig}
\usepackage{latexsym}
\usepackage{bm}

\def\beqr{\begin{eqnarray}}
\def\eqnr{\end{eqnarray}}
\def\beq{\begin{equation}}
\def\bc{\begin{center}}
\def\ec{\end{center}}
\def\eqn{\end{equation}}

\def\prl#1#2#3{{ Phys. Rev. Lett.} {\bf #1}, #2 (#3)}
\def\epl#1#2#3{{ Euro. Phys. Lett.} {\bf #1}, #2 (#3)}

\def\pre#1#2#3{Phys. Rev. E {\bf #1}, #2 (#3)}

\def\pnas#1#2#3{Proc. Natl. Acad. Sci. (USA) {\bf #1}, #2 (#3)}

\def\sc#1#2#3{Science {\bf #1}, #2 (#3)}

\def\sgn{\mbox{sgn}}

\begin{document}

\title{Scaling theory for the $1/f$ noise}

\author{Avinash Chand Yadav\footnote{jnu.avinash@gmail.com}}
\affiliation{Department of Physics, Institute of Science,  Banaras Hindu University, Varanasi 221 005, India}

\author{Naveen Kumar}
\affiliation{Department of Physics \& Astronomical Sciences, Central University of Jammu, Samba 181 143, India}

\begin{abstract}
{ We show that in a broad class of processes that show a  $1/f^{\alpha}$ spectrum, the power also explicitly depends on the characteristic time scale. Despite an enormous amount of work, this generic behavior remains so far overlooked and poorly understood.  An intriguing example is how the power spectrum of a simple random walk on a ring with $L$ sites shows $1/f^{3/2}$ not $1/f^2$ behavior in the frequency range  $1/L^2 \ll f \ll 1/2$.  
We address the fundamental issue by a scaling method and discuss a class of solvable processes covering physically relevant applications. }
\end{abstract}

\maketitle

The spectral content of a noisy process unveils the temporal correlation, a fundamental characteristic of dynamical systems. As a result, the spectral properties have been investigated at length, both experimentally and theoretically.
In particular, the scale-free power spectrum -- $1/f^{\alpha}$ form with the exponent $\alpha$ lying between 1 and 2 -- is termed {\it $1/f$ noise}~\cite{Dutta_1981, Weissman_1988, Montroll_1982, West_1989, Grinstein_1992, Milotti, Milotti_2005, Gleeson_2005, Eliazara_2009, Dahmen_2011, Amir_2012, Miguel_2014, Miguel_2015, Miguel_2018, Pereira_2019, Schwarz_2018, Krapf_2019}. Such a noise shows the existence of long time-correlation, a sign of complexity observed in nature, and one can find the exemplary instances in diverse contexts. In electronics, the examples include current or voltage fluctuations in vacuum tubes~\cite{Johnson_1925} and other devices. In life sciences, these range from neuronal activity, even at single neuron level~\cite{Siwy_2002, Lee_2005, Pettersen_2014}, to DNA sequence~\cite{Kaneko_1992, Voss_1992}. Other instances span from price fluctuations in economics~\cite{Stanley_1999} to online social activity~\cite{Jensen_2013}.  
In dynamical systems, the $1/f$ spectra emerge in intermittent chaotic systems~\cite{Benmizrachi_1985, Geisel_1987, Kaneko_2018} and energy spectra in quantum chaotic systems~\cite{Retamosa_2004, Kanzieper_2017}.

Given the wide occurrence,  the existence of a simple and general explanation might be suggestive. However, continued efforts to understand the noise have uncovered a few common explanations.  (i) The most widely applicable proposal is treating the process as a superposition of several independent exponentially relaxing events, with a power-law relaxation time distribution. (ii) The hypothesis of self-organized criticality (SOC)~\cite{Bak_1987, Bak_1996, Maslov_1999, Zhang_1999, Davidsen_2002, Laurson_2005, Dhar_2006, Yadav_2012} explains scale-invariant features, both in space and time, for a class of non-equilibrium systems. (iii) A class of nonlinear stochastic differential equations with multiplicative noise can also generate the $1/f$ noise~\cite{Kaulakys_2005, Ruseckas_2010, Ruseckas_2011, Ruseckas_2014, Ruseckas_2016}. (iv) A memoryless nonlinear response (MNR) serves as an alternative mechanism. A noisy stimulus, instantaneously processed through nonlinear transformation, results in a response that may show the $1/f$ noise~\cite{Yadav_2013, Yadav_2017}.

\begin{figure}[t]
  \centering
  \scalebox{0.3}{\includegraphics{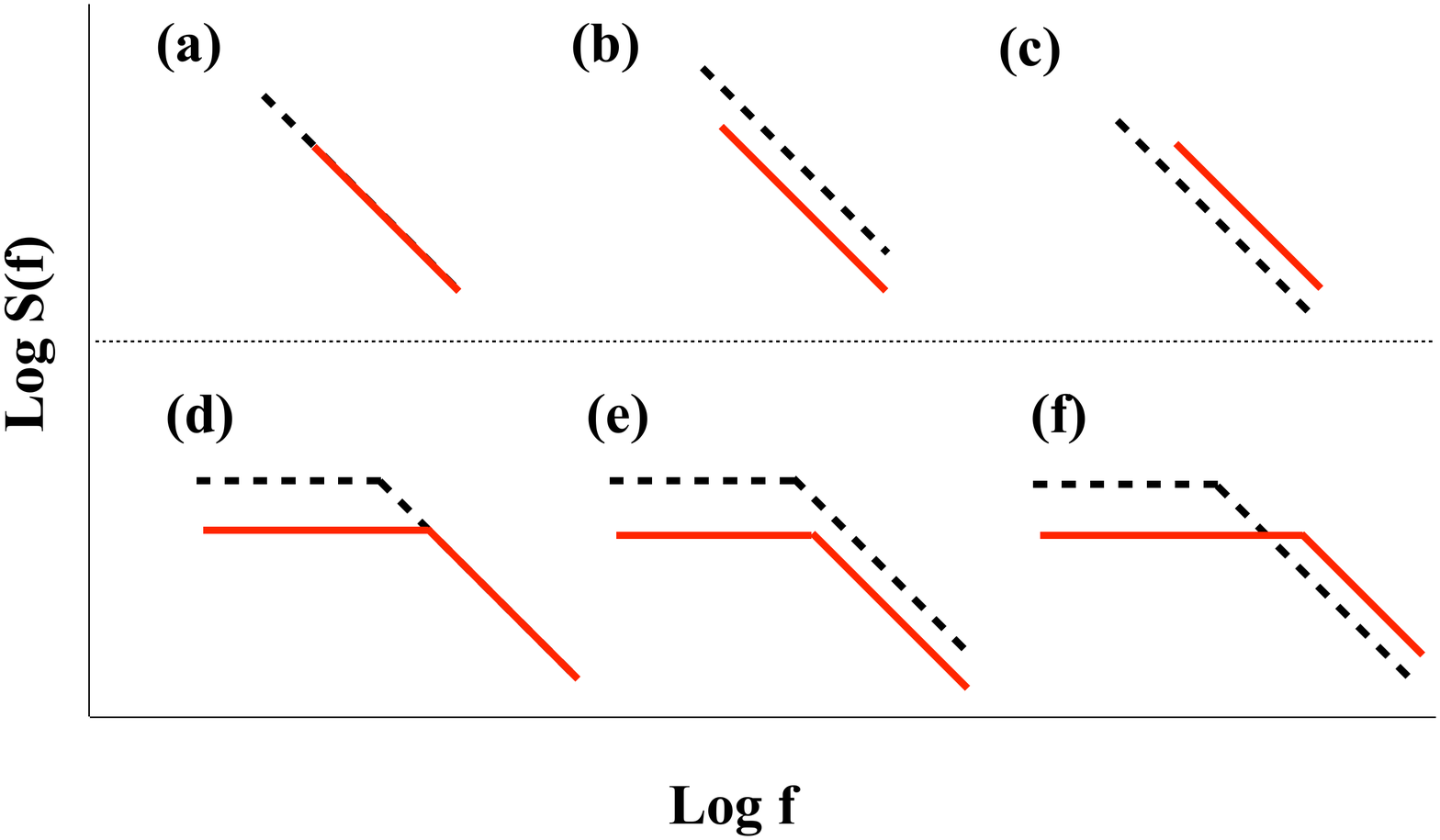}}
  \caption{{Schematic diagram for possible scaling features of power spectrum showing $1/f$ type form, along with dependence on finite observation time [(a) to (c)] or characteristic time scales [(d) to (f)]. The solid and dashed curves correspond to two different values of the time.}}
  \label{fig_sch_diag}
\end{figure}

Although this much is  generally agreed and understood, the $1/f$ noise continues to be reasonably active because of its enigmatic nature. In recent years, the focus has been on various aspects. One pertinent aspect is the ``aging'' feature. The power spectrum also depends on the finite observation time of the process [see Fig.~\ref{fig_sch_diag} (a) to (c)]. Firstly, it is theoretically suggested for non-stationary processes~\cite{Mandelbrot_1967},  which later experimentally observed in blinking quantum dots~\cite{Niemann_2013, Sadegh_2014, Leibovich_2015}, current fluctuations in nanoscale electrodes~\cite{Krapf_2013}, and interface fluctuations in liquid-crystal turbulence~\cite{Kazumasa_2017}.

 \begin{figure*}[t]
  \centering
  \scalebox{0.485}{\includegraphics{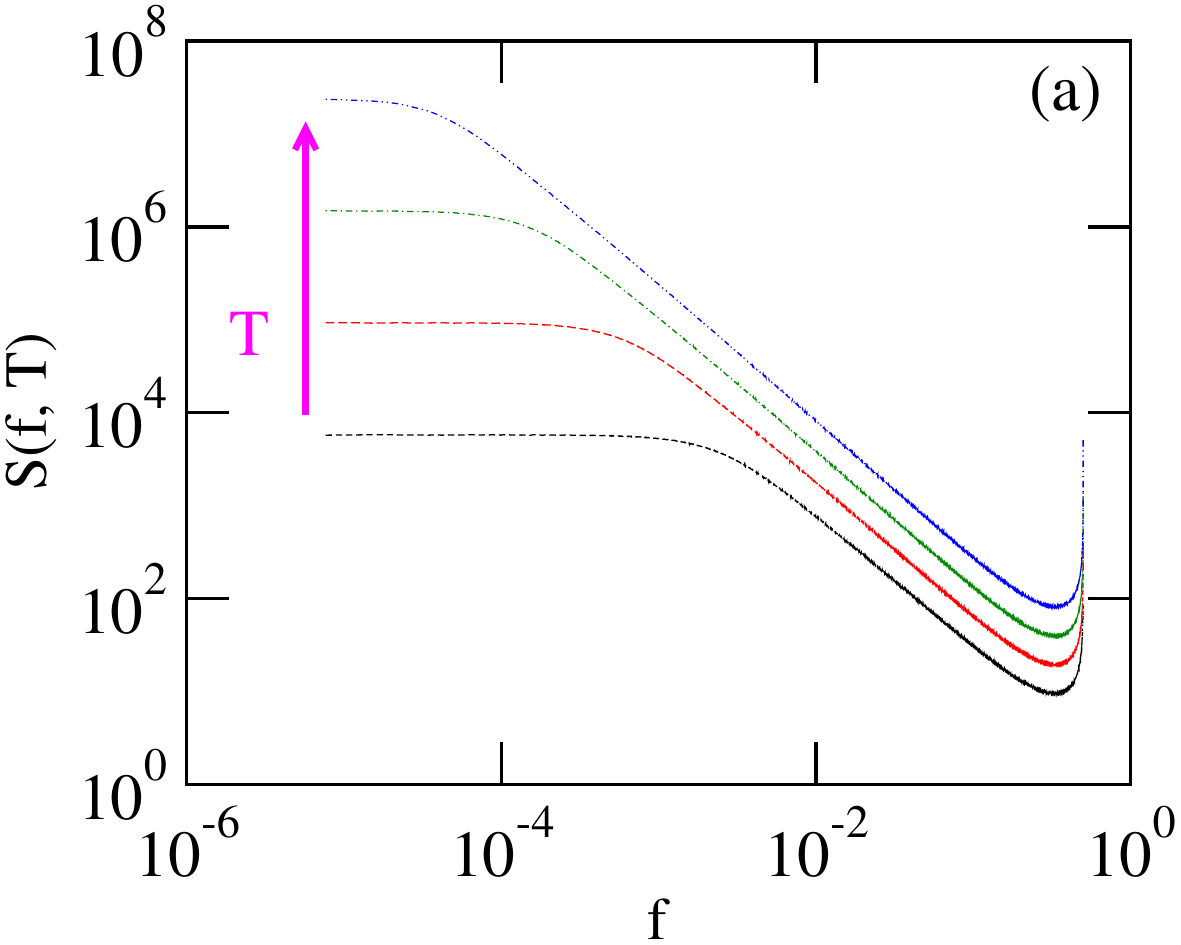}}
  \scalebox{0.485}{\includegraphics{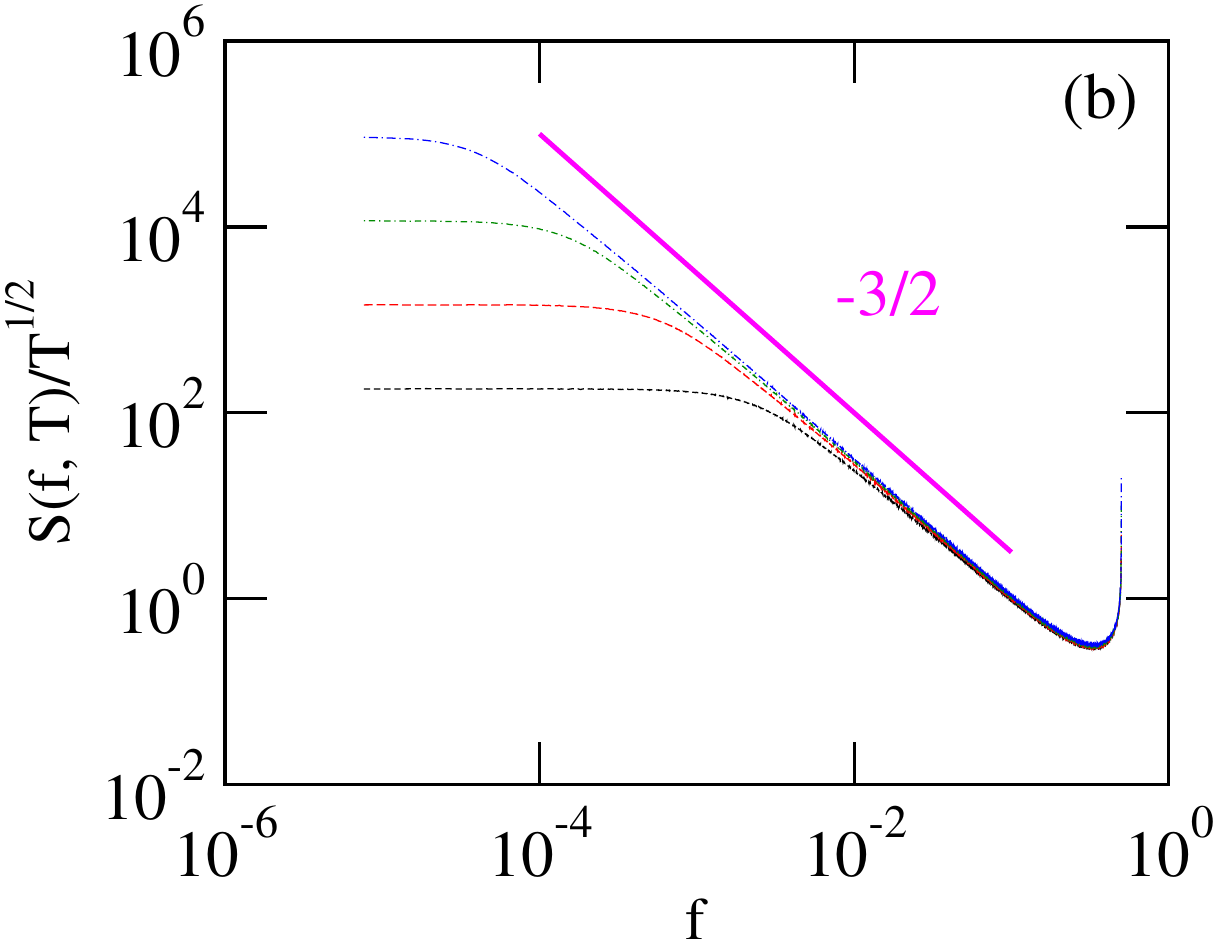}}
  \scalebox{0.485}{\includegraphics{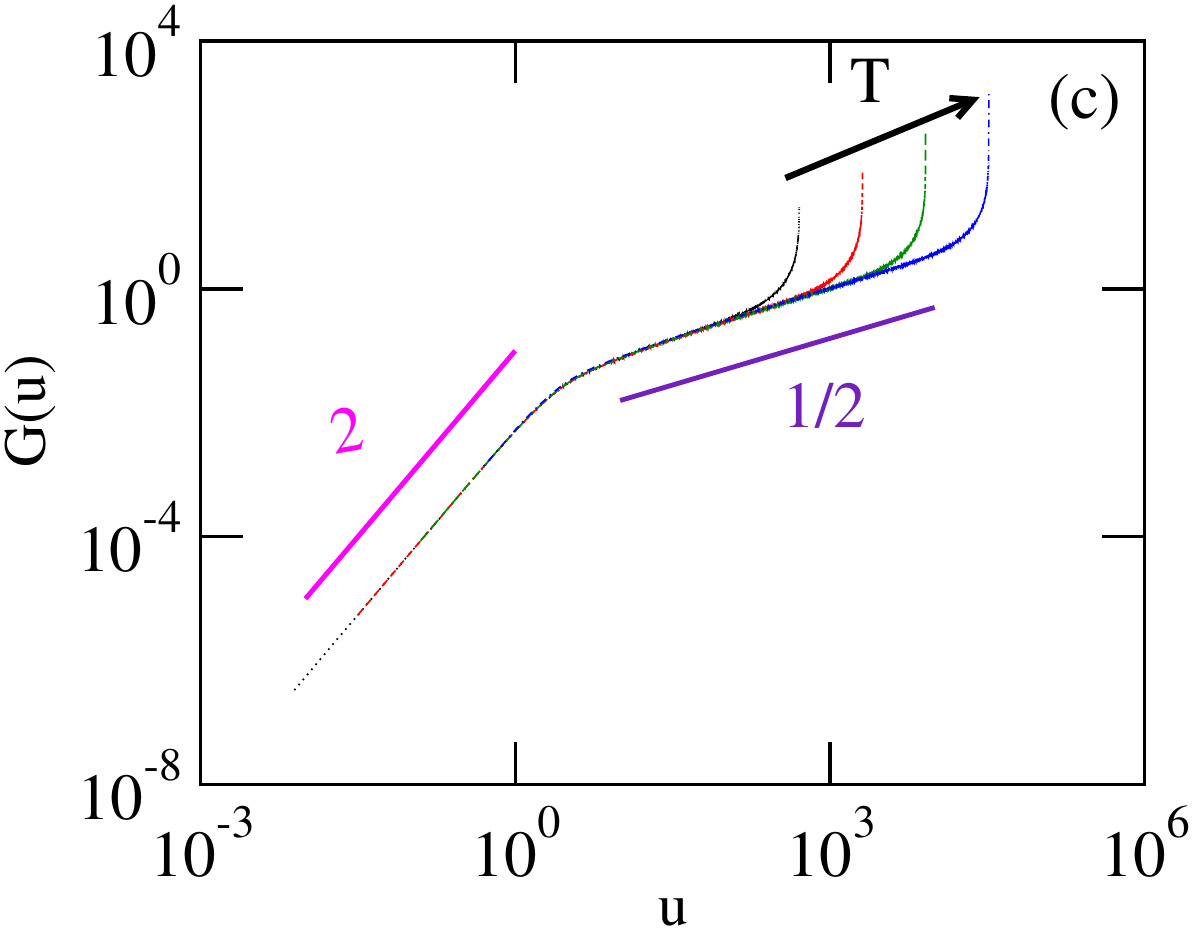}}
  \caption{(a) The power spectra for the SRW on a ring of size $L$.  Here, the time series length is $N = 2^{18}$ and averaging is performed  over  $10^4$ ensembles. The arrow indicates  trend of the curves with  different increasing values of $T = 2^{10},  2^{12}, 2^{14}$, and  $2^{16}$.     (b) Quantification of the aging feature: The data collapse  for the curves shown in (a).  (c) The scaling function: $\mathcal{G}(u) = f^2 S_{\xi}(f,T)$ with $u = fT$. The straight lines with respective slopes are drawn for a comparison.}
  \label{fig_2}
\end{figure*}

In this paper, we show a generic feature for a broad class of the $1/f$ noise: The power also explicitly depends on the {\it characteristic time scale} [see Fig.~\ref{fig_sch_diag} (e) and (f)]. One can note the behavior in sandpile \cite{Zhang_1999} and a class of $1/f$ processes found within the MNR framework~\cite{Yadav_2017}.  Despite much research, the unusual behavior remains so far overlooked and poorly understood from the aspect of origin and physical implications it can have.  
We take the first step to address the fundamental problem by a scaling theory to describe this dependence on the characteristic time scales or the length of the time series of observation, for processes that are not stationary in time. The scaling method is crucial in the critical phenomena and to the systems displaying scale-invariant properties~\cite{Christensen_2005, Djordje_2011, Djordje_2011_pre, Manchanda_2013, Avinash_2018}. We explain below the simplest non-trivial example and then suggest a general framework with striking instances.

Consider the following scenario. A noisy sequence $x(t)$, taking values 1 or -1 with equal probability, represents white noise with the power spectrum $\sim 1/f^0$. A linear operation such as integration of white noise  generates Brownian noise $\sim 1/f^2$, typically modeled as a simple random walk (SRW) on a line: $\xi(t) = \sum_{i=1}^{t}x(i)$. The unbounded process is non-stationary, and the power spectrum does not show explicit dependence on the finite observation time $T^*$.  If we impose a reflecting boundary conditions at $\xi = \pm L$, this introduces a characteristic time scale $T \sim L^2$. The power spectrum remains unchanged, except showing $T$ dependence similar to Fig.~\ref{fig_sch_diag} (d). Note that the bounded normal diffusion results in a time scale $T \sim L^2$. As the $T$ increases, the total power of the signal increases linearly. Thus, the power spectrum scales as $1/f^2$ for $1/L^2 \ll f \ll 1/2$.

We can now present the simplest non-trivial example of the $1/f$ noise with explicit aging as a function of characteristic time. If we impose a periodic boundary to the SRW, then the walker runs on a ring of size $L$ with $\xi \in [0, L-1]$ such that $T = L^2$. 
Such a boundary condition produces a nonlinear effect that results in explicit characteristic time dependence, and eventually, a non-trivial behavior emerges for the power spectrum as $\sim 1/f^{3/2}$   for $1/T \ll f \ll 1/2$ [see Fig.~\ref{fig_2}]. It is easy to note that the total power remains $\sim T$. The typical magnitude of the process $L \sim \sqrt{T}$ appears in the picture.
Noting the probability to find the walker at $\xi$ that is $\mathcal{P}(\xi, T) = 1/L = 1/\sqrt{T}$ for $0 \le \xi \le L-1$, we find the total power
$ P_{\xi}(T) = \langle \xi^2 \rangle  =  \int_{0}^{\sqrt{T}} \xi^2  d\xi/\sqrt{T}    \sim T.$ One may naively expect $1/f^2$ type behavior that is contrary to the actual behavior. While a Ref.~\cite{Erland_2007} suggests $1/f^{3/2}$ feature for cyclical random walk, it does not discuss the explicit characteristic time dependent feature.  We show that the scaling methods can offer a simple explanation for this non-trivial observation.

The power at a fixed frequency increases with the characteristic time over the entire range of frequencies, but the rate differs in the two frequency regimes. 
From the numerical results shown in Fig.~\ref{fig_2}, one can readily write an expression for the power spectrum
\begin{equation}
S_{\xi}(f, T) = \begin{cases}A T^2, ~~~~~~~ {\rm for}~~ f\ll \frac{1}{T}\\ A\frac{\sqrt{T}}{f^{3/2}},~~~~~{\rm for~~} \frac{1}{T} \ll f \ll \frac{1}{2}.\end{cases}
\end{equation} 
The form does not capture the exact behavior around the frequency  $1/T$. Since the power spectrum $S_{\xi}(f, T)$ is a homogeneous function of its arguments, this can be re-expressed as
\begin{equation}
S_{\xi}(f, T) = A\frac{1}{f^2} \mathcal{G}(fT).
\end{equation}
The scaling function is $ \mathcal{G}(fT) \sim (fT)^{2}$  for $f\ll 1/T$ and
$\mathcal{G}(f T) \sim  \sqrt{fT}$ for  $1/T \ll f \ll 1/2$.
Clearly, the scaling function behaves as
\begin{equation}
 \mathcal{G}(u) \sim \begin{cases} u^2,~~~~~{\rm for~~~}u\ll1, \\ \sqrt{u},~~~~ {\rm for~~~}u\gg 1,\end{cases}
\end{equation}
where $u=fT$.
The total power content of the signal is
\begin{equation}
P_{\xi}(T) =  \int_{0}^{\infty} S_{\xi}(f, T) df  = AT \int_{0}^{\infty}\frac{du}{u^2}\mathcal{G}(u)  \sim T.\nonumber
\end{equation}

In the example, the {\it boundary constraint} is the attribute responsible for the origin of the $1/f$ noise with explicit $T$ dependence. Note that MNR is an alternative mechanism that can explain such explicit aging and the subsequent emergence of the $1/f$ noise.

We point out an interesting example where the process arises naturally. The noisy process $\xi(t)$, a trajectory of the SRW on a ring, appears in a directed Abelian sandpile model that displays SOC~\cite{Maslov_1999, Yadav_2012}. The process $\xi$ denotes the evolution of the configuration of a sandpile. 
Since the sandpile model is important for our discussion, we here briefly describe this. Consider a $2\times n$ lattice, where $n$ is the linear extent. The number of allowed configurations are $3^n$, since each column $2\times i$ can take one of the possible recurrent states  $\zeta_i \in \{
\big(\begin{smallmatrix}
  0\\
  1
\end{smallmatrix}\big),
\big(\begin{smallmatrix}
  1\\
  1
\end{smallmatrix}\big),
\big(\begin{smallmatrix}
  1\\
  0
\end{smallmatrix}\big)\}$. 
The system is driven by adding one particle at the top or bottom site at one end, and the particles can leave the system from the other end. The driving may cause an unstable configuration, having sites with 2 or more particles, and the system relaxes by a simple toppling rule. An unstable top or bottom site transfers one particle each to the right and bottom or top sites. Taking $\zeta_i \in \{-1, 0, 1\}$ as a bit in ternary-base, one can relate each stable configuration to an integer. In turn, the complicated dynamics reduce to a simple random walk on a ring with $ 3^n$ sites. Counting zeros in the ternary base for $\xi$, one can determine the total mass of the system $\eta$ that shows $1/f$ noise.  Mathematically, $\eta = n + \sum_i q_i$, where $q_i = 1$ if $\zeta_i = 0$ and $q_i = 0$ if $\zeta_i \ne 0$.  Note that the indicator function $q_i$ captures the local mass fluctuations, which we discuss later.

{\bf \it General Formalism.--}
We can recast the analysis of the above example in a more general manner. We begin with processes, where the power spectrum shows aging as a function of characteristic time scale. Mathematically, the typical power spectrum reads 
\begin{equation}
S(f,T) = \begin{cases}A T^{\alpha'+\beta'},~~~{\rm for}~~f\ll 1/T \\AT^{\beta'} \frac{1}{f^{\alpha'}},~~~{\rm for}~~1/T\ll f\ll 1/2,\end{cases}
\end{equation}
where $\alpha'$ and $\beta'$ are spectral and aging exponents, respectively. The characteristic time scale $T$ is a system-specific property, and it may appear because of finite system size. The aging feature generally appears in two forms. (i) The power spectrum exhibits time-dependent behavior only in the regime $f\ll 1/T$ if $\beta' = 0$. (ii) If $\beta' \ne 0$, the power spectrum depends on $T$ even in the non-trivial regime $f\gg 1/T$. We call this case ``explicit aging''.

However, in some experimental studies, no characteristic time scale has been observed~\cite{Niemann_2013}. Note that the total power for $1/f$ noise would be infinite if there were no characteristic time scale. Interestingly, the spectral content shows aging as a function of the finite observation time $T^*$. Examples include single file diffusion and Brownian motion in logarithmic potential~\cite{Leibovich_2015}; see Ref.~\cite{Dechant_2015} for more instances. It is a class of non-stationary $1/f$ processes, where the regime $f\ll 1/T^*$ does not exist.  Also, the finite observation time dependence for $1/f$ noise remains well recognized. In this work, we mainly focus on $1/f$ processes showing explicit characteristic time scale dependence.

We consider a noisy time series $\xi (t)$, with a temporarily scale-invariant feature. A relevant quantity is the total power $P = \langle \xi^2\rangle$, where the angular bracket $\langle \cdot \rangle$ denotes ensemble average. Typically, the total power diverges as a function of the characteristic time scale $T$. We focus on a class of processes for which $P(T) \sim T^{\gamma}$, with $0 \le \gamma \le 1$. Note that Parseval's theorem relates the total power of the process with the power spectrum $P(T) =   \int S(f,T)df$.   Recognizing scaled frequency $u = fT$, the frequency dependency of the power spectrum can be determined
\begin{equation}
S(f) \sim \left |\frac{dP[T(f)]}{df}\right | \sim \frac{1}{f^{\alpha}}, ~~~{\rm for}~~\frac{1}{T} \ll f \ll \frac{1}{2},\nonumber
\end{equation}  
with $\alpha = 1+\gamma$.
In general $\alpha \ne \alpha'$.

Recall that the power spectrum shows aging, $T$ dependent feature. 
It implies that the power spectrum is a function of both variables $f$ and $T$. The scaling feature suggests that the power spectrum $S(f, T)$ is a homogenous function of its arguments. Then, the power spectrum can be easily expressed as
\begin{equation}
S(f, T) = S(f)\mathcal{G}\left(fT\right) = \frac{1}{f^{\alpha}} \mathcal{G}(u) = T^{\alpha}\mathcal{H}(u),
\label{eq_scaling}
\end{equation} 
where $u = fT$.
Asymptotically, the scaling function varies  as
\begin{equation}
\mathcal{G}(u) = \begin{cases} \mathcal{G}_<(u) \sim u^{\nu'},~~~~~{\rm for}~~~~u\ll 1,\\ \mathcal{G}_>(u)\sim u^{\nu},~~~~~~{\rm for}~~~~u\gg 1.\end{cases}
\end{equation}
Similarly, $\mathcal{H}_{<}(u) \sim $ constant and $ \mathcal{H}_{>}(u) \sim 1/u^{\alpha'}$.
Moreover, if the power spectrum is frequency independent $S(f,T) = S(f = 1/T)$  for $f\ll 1/T$, then $\nu'  = \alpha =  1+\gamma$. Since $\langle \xi^2\rangle \sim T^{\gamma}$, the typical magnitude  of the process is $L \sim \sqrt{T^{\gamma}}$. In many cases, the typical aging is found to be directly or inversely proportional to $L$ in the regime $f\gg 1/T$. Then, $\nu = \pm \gamma/2$.

What do we directly learn from the scaling function, numerically obtained as data collapse of the power spectra? First, we can estimate the exponent of the total power from the scaling function exponent in the regime $u\ll 1$,  as $\gamma = \nu'-1$. Second, the scaling exponent quantifies the aging $\beta' = \nu$ for $u\gg 1$. If $\nu \ne 0$, the explicit aging occurs. The sign of the exponent $\beta'$ is $\pm$ if the power spectrum is directly or inversely related to $T$. Eventually, the two exponents alone suffices to estimate spectral characteristic exponents:  $\alpha' = \nu'-\nu$, $\beta' = \nu$, $\gamma = \nu'-1$, and $\alpha = \nu'$. Thus, the scaling function can especially offer useful insight where theoretical progress remains challenging.

\begin{figure}[t]
  \centering
  \scalebox{0.56}{\includegraphics{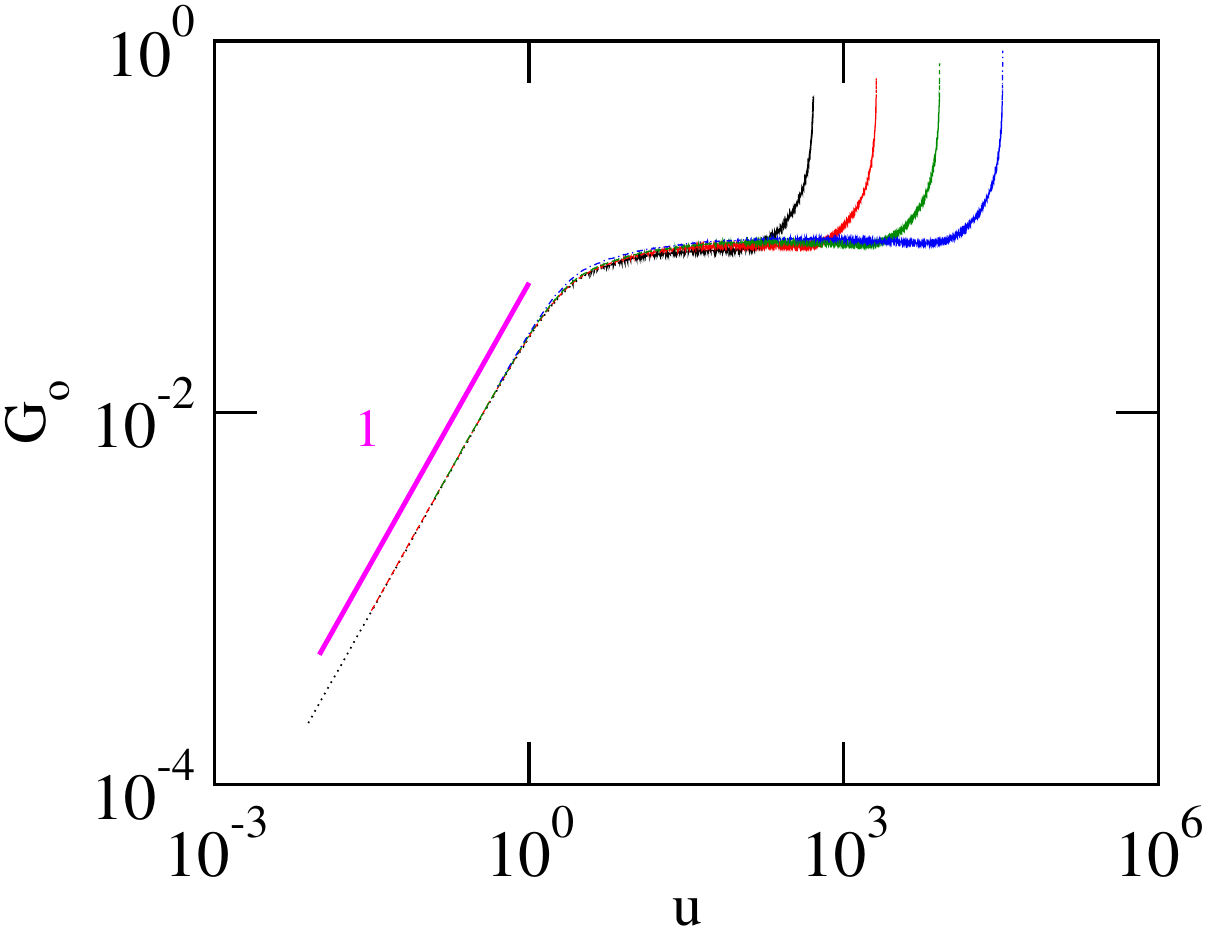}}
    \caption{ The scaling function for a nonlinear response functions $\eta = \mathcal{R}(\xi)$,  with an input noise shown in Fig.~(\ref{fig_2}).  $\mathcal{R}(\xi)$ is the sum of 1's in binary expansion of $\xi$. }
    \label{fig_3}
\end{figure}

{\bf \it Effect of nonlinearity.--} To examine the extent of the scaling theory, we present examples of analytically tractable processes. Nonlinear response to noisy stimuli can model many processes. 
One example is the total mass fluctuations in the sandpile model, exhibiting the $1/f$ spectrum. The nonlinear response $\eta$   is the count of zeros in binary or ternary base for $\xi$~\cite{Yadav_2012}. In the second example, we consider the response of a sensory system to a noisy stimulus. Here, the response function varies in a sub-linear manner. From psychophysics, we well know two phenomenological instances:  Weber-Fechner law ($\eta \sim \log \xi$) and Stevens' law ($\eta \sim \xi^a$)~\cite{Stevens_1975, Copelli_2002}. Assuming the nonlinear response to be memoryless, we explore spectral characteristics for these physically relevant transformations with input noise modeled as $\xi(t)$. Figure~\ref{fig_3} shows numerical results for the scaling function with a nonlinear transfer function discussed below [see Eq.~(\ref{eq_s1b})].

 \begin{figure}[t]
  \centering
  \scalebox{0.56}{\includegraphics{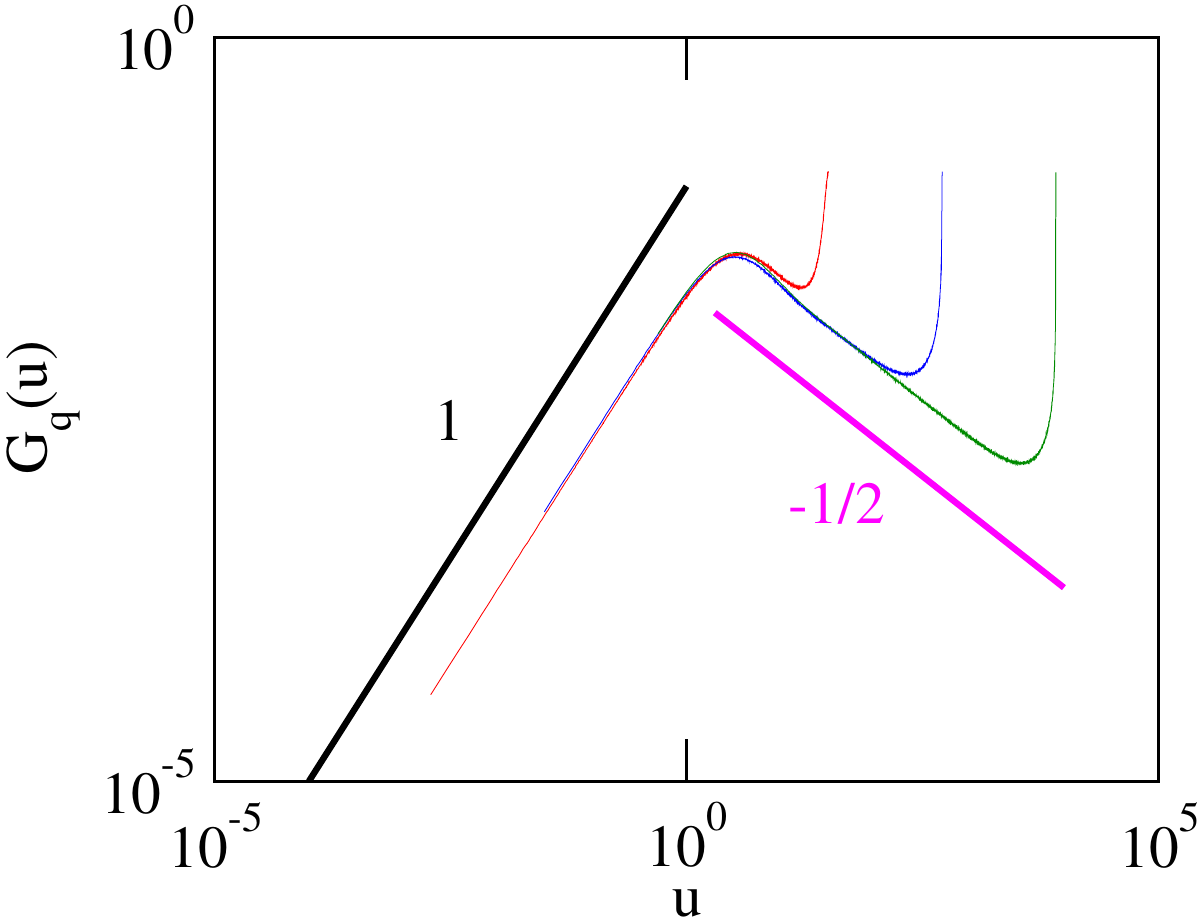}} 
  \caption{The scaling function for fluctuations in $q_i$ for $i = 3, 5$ and 7 with $n = 10$. Here, the characteristic time is $T(i) = 2^{2i}$. }
    \label{fig_4}
\end{figure}

\begin{itemize}
\item For $\eta$ to be sum of 1's in binary expansion of $\xi$, the power spectrum varies as $\sim 1/f$. We note
\begin{subequations}
\begin{align}
S_{\eta}(f, T) =  A_o\frac{1}{f} \mathcal{G}_o(fT),
 \label{eq_s1a}
\end{align}
where
\begin{align}
 \mathcal{G}_{<}(u) \sim u,~~~~~{\rm and~~~~~~} \mathcal{G}_{>}(u) \sim  1.
 \label{eq_s1b}
 \end{align}
\end{subequations}

\item We also examine the power spectrum for the local mass fluctuations of the sandpile $q_i$ and note $1/f^{3/2}$ behavior along with $1/\sqrt{T}$ dependence in the regime $f\gg1/T$ [see Fig.~\ref{fig_4}].  We have 
\begin{subequations}
\begin{align}
S_{q_i}(f, T) =  A_o\frac{1}{\sqrt{T}f^{3/2}} = A_o \frac{1}{f}\mathcal{G}_q(fT),
 \label{eq_ls1a}
\end{align}
where
\begin{align}
 \mathcal{G}_{<}(u) \sim u,~~~~~{\rm and~~~~~~} \mathcal{G}_{>}(u) \sim  \frac{1}{\sqrt{u}}.
  \label{eq_ls1b}
 \end{align}
\end{subequations}

\end{itemize}

It is easy to note that the $1/f$ power spectrum of the total mass in the sandpile is equal to the sum of local $1/f$ power spectra [see Eqs.~(\ref{eq_s1a}) and (\ref{eq_ls1a})] since the local fluctuations are uncorrelated in space.

Within the memoryless nonlinear response framework,  consider a sublinearly varying transfer function $\eta = \sgn(\xi)|\xi|^a$ with $0< a \le 1/2$. In this case, the spectral properties have been recently computed, analytically, with two classes of noisy input, namely, Brownian and Gaussian process with spectral feature varying as $1/f^{1+b}$~\cite{Yadav_2013, Yadav_2017}. To gain a better understanding of the $T$ dependent spectral behavior, we re-examine focusing on the scaling function.

\begin{itemize}
\item When we model the input as a SRW with reflecting boundary at $\xi = \pm L$, the power spectrum of the output varies for the frequency $f\gg 1/T$  as 
\begin{subequations}
\begin{align}
S_{\eta}(f, T) = A_o \frac{1}{f^{1+a}}\frac{1}{\sqrt{fT}} =  A_o\frac{1}{f^{1+a}}\mathcal{G}_o(fT),
\end{align}
where 
\begin{align}
 \mathcal{G}_{<}(u) \sim u^{1+a},~~~~~{\rm and~~~~~~} \mathcal{G}_{>}(u) \sim \frac{1}{\sqrt{u}}.
 \label{eq_bn_rb}
\end{align}
\end{subequations}
Here, the total power is $P_{\eta} \sim T^a$.

\item If the input  is a Gaussian process displaying a power spectrum of $1/f^{1 + b}$ type for $f\gg 1/T$ with $0\le b \le 1$, the power spectrum of the output process behaves as 
\begin{subequations}
\begin{align}
S_{\eta}(f, T) =  A_o\frac{1}{f^{1+a b}}\frac{1}{\sqrt{(fT)^b}} =A_o\frac{1}{f^{1+a b}}\mathcal{G}_o(fT),
\end{align}
where 
\begin{align}
 \mathcal{G}_{<}(u) \sim u^{1+a b},~~~~~{\rm and~~~~~~} \mathcal{G}_{>}(u) \sim \frac{1}{\sqrt{u^b}}.
 \label{eq_gn}
\end{align}
\end{subequations}
In this case, the total power varies as $P_{\eta} \sim T^{ab}$.
\end{itemize}

\begin{figure}[t]
  \centering
  \scalebox{0.56}{\includegraphics{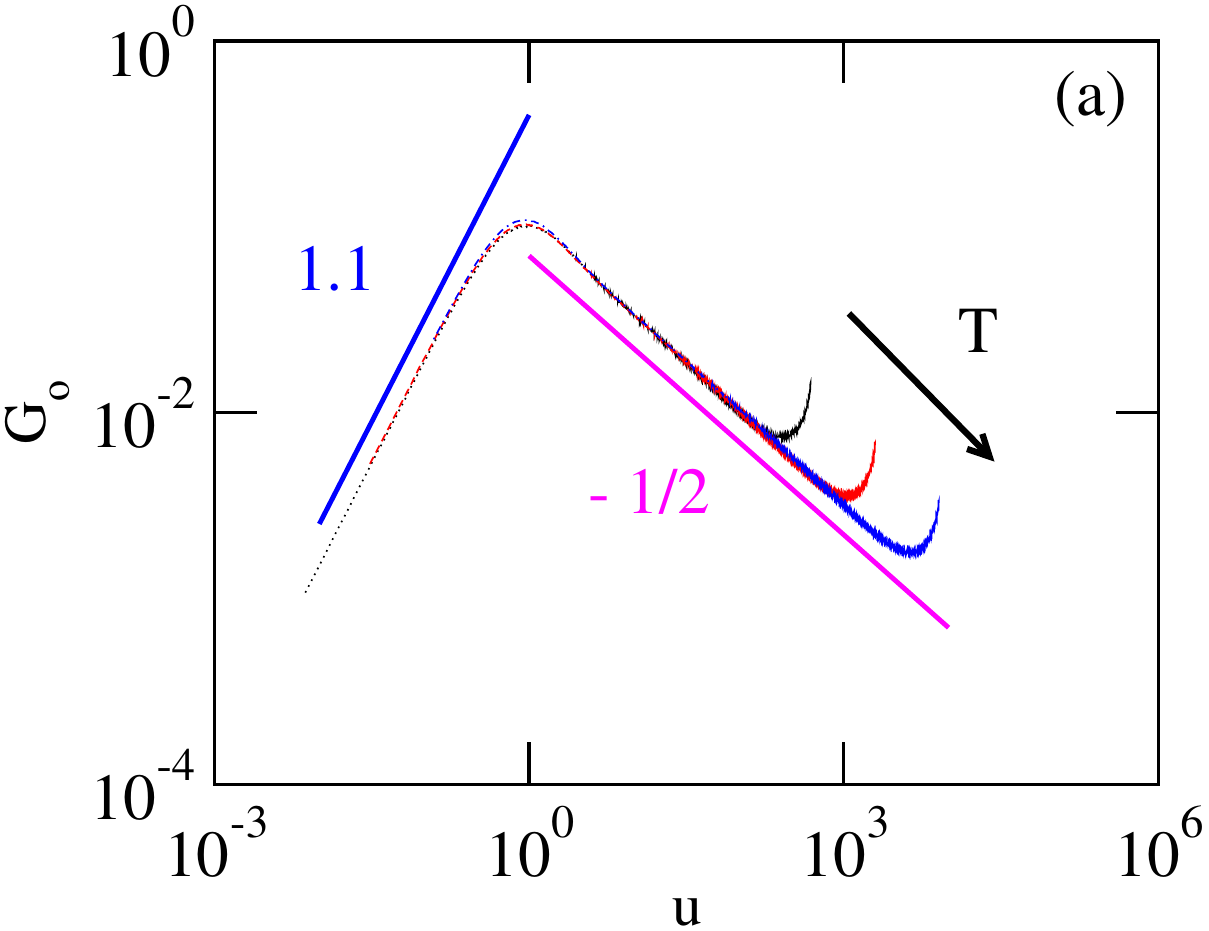}}
 \scalebox{0.56}{\includegraphics{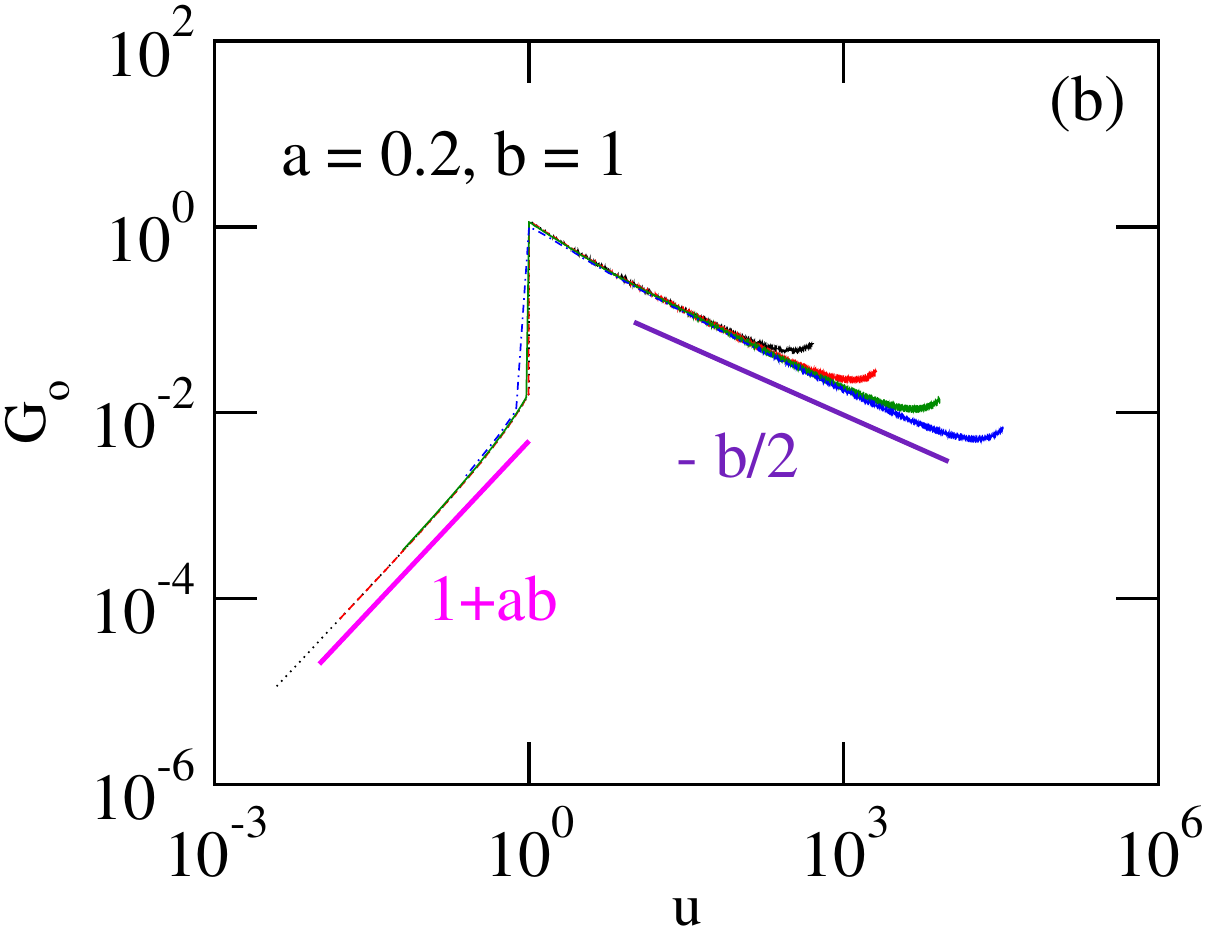}}
 \caption{{ The scaling function for a response function  $\eta = \sgn(\xi)|\xi|^a$ with different noisy inputs.} (a) The input is modelled as SRW on line with a reflecting boundary at $\xi = \pm L $.  Here, $a = 0.1$ and $T = (2L-1)^2$, with $L = 2^4, 2^5,$ and $2^6$. (b) The input is a  Gaussian process with $\sim 1/f^{1 + b}$ for $b = 1$ and $T = 2^{10}, 2^{12}, 2^{14}$, and $2^{16}$. }
    \label{fig_5}
\end{figure}

Numerical results shown in Fig.~\ref{fig_5} provide a validation for theoretically predicted behavior described by Eqs.~(\ref{eq_bn_rb}) and (\ref{eq_gn}). An excellent agreement is seen with the theory within statistical error.
We briefly outline implemented numerical methods. We generate the input processes using the Monte Carlo method. See the Ref.~\cite{Yadav_2017} for the Gaussian process with spectral property $1/f^{1+b}$ for $f\gg 1/T$. The output is easy to compute by applying instantaneous nonlinear transformation upon the input. We implement the fast Fourier transform algorithm to compute the power spectrum of the response. Since the power spectrum of a single realization shows a highly fluctuating behavior, we employ ensemble averaging to get a smooth curve. Only the frequency regime $f\ll 1/2$ is of interest. As our primary focus is on data collapse, this is easy to compute as $\mathcal{G}(u) = f^{\alpha}S(f, T)$ with $u = fT$. To numerically get the data collapse, the exponent $\alpha$ and the time scale $T$ need to be determined. A theoretical estimate of the total power can provide the exponent $\alpha$.

A few observations are in order. 
Beyond the spectral exponents, the scaling function reveals the unique nature of the underlying process. As the different spectral exponents correspond to a distinct universality class, the data collapse offers an alternative to estimating spectral exponents.
The non-trivial spectral exponent also arises by altering the boundary constraint, and the explicit time dependency of the power spectrum may appear or disappear because of nonlinearity. For the $1/f$ noise with $\alpha = 1$, no explicit aging occurs; the implication is consistent with a similar remark noted in the Ref.~\cite{Gerardo_2010}. Our framework is also consistent with well-recognized $1/f$ processes showing explicit finite observation time dependence.

In summary, we have introduced a scaling theory for the $1/f$ noise,  explaining the associated aging feature and its consequences. In particular, we offer an adequate insight for the $1/f$ noise with an explicit characteristic time-dependent feature.  
We consistently apply the theory to a class of spatially extended dynamical systems and recognize the explicit aging for $1/f$ noise~\cite{avinash_tm}. It is easy to analyze experimentally collected noisy signals.  The scaling approach for the $1/f$ noise is simple and of broad relevance due to its general applicability.

ACY warmly acknowledges support from a grant ECR/2017/001702 funded by SERB, DST, Government of India. NK would like to acknowledge financial support from the CUJ-UGC fellowship.  ACY would also like to deeply acknowledge Deepak Dhar and Ramakrishnan Ramaswamy for careful reading the manuscript and providing useful suggestions.


\begin{thebibliography}{99}
\bibitem{avinash_tm} The results will be communicated in a separate paper.
\end{thebibliography}
\end{document}